\documentclass[letterpaper, 10 pt, conference]{ieeeconf}  % Comment this line out
\IEEEoverridecommandlockouts                              % This command is only
\usepackage[utf8]{inputenc}
\usepackage[T1]{fontenc}
\usepackage{hyperref}
\usepackage{cleveref}
\usepackage{graphicx} %LaTeX package to import graphics
\graphicspath{{images/}} %configuring the graphicx packag
\usepackage{multirow}
\usepackage{booktabs}
\usepackage{siunitx}

\title{\LARGE \bf
Bayesian Optimization for Adaptive Fraud Authentication in \\Interactive Voice Response System
}

\author{Jingrong Xie \\ 
    \texttt{jennaxie@gatech.edu} \\ 
    \and 
    Yuming Li \\ 
    \texttt{liyumin@udel.edu} \\
    }
\begin{document}

\maketitle
\thispagestyle{empty}
\pagestyle{empty}

%%%%%%%%%%%%%%%%%%%%%%%%%%%%%%%%%%%%%%%%%%%%%%%%%%%%%%%%%%%%%%%%%%%%%%%%%%%%%%%%
\begin{abstract}

This paper introduces a Bayesian approach to improve Interactive Voice Response (IVR) authentication processes used by financial institutions. Traditional IVR systems authenticate users through a static sequence of credentials, assuming uniform effectiveness among them. However, fraudsters exploit this predictability, selectively bypassing strong credentials. This study applies Bayes' Theorem and conditional probability modeling to evaluate fraud risk dynamically and adapt credential verification paths. Through simulation experiments using real-world-inspired data, we develop algorithms to identify the most effective credential combinations given certain conditions and propose dynamic selection from available credentials. The findings suggest an optimized, adaptive authentication flow that balances fraud detection with user convenience, providing a road map for banks to enhance security within automated channels.

\end{abstract}

%%%%%%%%%%%%%%%%%%%%%%%%%%%%%%%%%%%%%%%%%%%%%%%%%%%%%%%%%%%%%%%%%%%%%%%%%%%%%%%%
\section{Background}

Since the onset of the COVID-19 pandemic, financial services and insurance companies have experienced a dramatic increase in contact center call volumes. Shifts in consumer behavior led customers to reach out in record numbers, with inquiries ranging from basic account maintenance to urgent financial needs. To manage this surge efficiently, organizations increasingly turned to Interactive Voice Response (IVR) systems \cite{Foreester}. 

IVR technology automates routine customer interactions, allowing callers to navigate a bank's phone system using either voice commands or keypad inputs \cite{BankingSolution}. Through this channel, customers can perform common self-service tasks, such as checking account balances, transferring funds, reviewing recent transactions, or updating personal information, without waiting to speak with a live agent. By streamlining access to frequently used services, IVR reduces call wait times, lower operational costs, and improves customer convenience \cite{IVRBenefit}. 

While IVR adoption has delivered clear benefits in scalability and customer experience, it has also coincided with a surge in fraudulent activity. Fraudsters have increasingly exploited automated systems, highlighting the dual challenge of maintaining customer convenience while strengthening security measures.

\begin{figure}[h]
    \centering
    \includegraphics[width=0.95\linewidth]{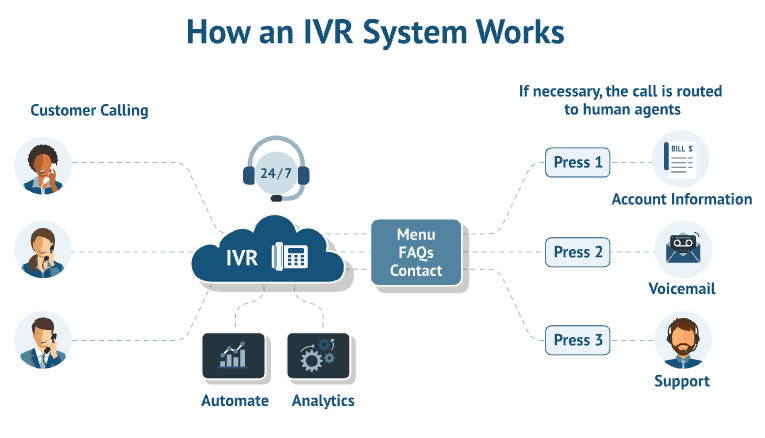}
    \caption{IVR System Flow Diagram}
    \label{fig:IVR_SYS_FLOW }
\end{figure}

To protect customer information, banks have integrated authentication mechanisms into IVR systems. These mechanisms often require callers to verify their identity using a combination of credentials such as credit card number, card verification value (CVV), social security number (SSN), Automatic Number Identification (ANI), ZIP code, or voice biometrics. While these measures are designed to ensure that only legitimate customers gain access, the security strength of these credentials varies. Publicly available information like ZIP code or address may be easier for fraudsters to obtain, while credentials such as CVV or voice biometrics are generally more secure. However, even stronger credentials may be compromised due to data breaches or social engineering \cite{Databreach}.

The original design of IVR systems treated all services accessed via IVR as low risk, assuming that passing authentication indicated a legitimate user. As a result, customers typically only needed to pass a fixed set of two or three credentials to be served within the IVR system. This design emphasized customer convenience, assuming uniform effectiveness of all credentials and treating successful authentication as a strong signal of legitimacy.

\section{Research Motivation}

Despite these security controls, fraudsters have increasingly exploited IVR systems to gather account information and identify transaction patterns. By repeatedly calling the IVR, testing credentials, and learning the structure of the authentication process, fraudsters are able to incrementally collect enough information to commit downstream fraud, such as unauthorized online transactions or fund transfers.

Current IVR authentication methods are static and rule-based, assuming all credentials offer equal security and that a successful authentication implies trust. However, real-world behavior tells a different story: fraudsters and legitimate customers exhibit quite different interaction patterns. Fraudsters tend to call from unknown or mismatched numbers, make repeated call attempts, and probe multiple accounts within short time frames. In contrast, genuine customers usually call from known numbers and interact only with their own accounts.

Moreover, fraudsters exploit the fixed order and predictability of IVR authentication, often bypassing stronger credentials and focusing on weaker ones. At the same time, legitimate customers may occasionally fail a strong credential due to forgotten PINs or input errors, which introduces additional uncertainty. These nuances highlight the need for a more adaptive and intelligent authentication process—one that can distinguish fraudsters from real customers with higher accuracy.

Recent work highlights the importance of capturing sequential context in behavior modeling, particularly in domains like fraud detection where isolated features often fail to represent underlying intent. Kawawa-Beaudan et al. \cite{Sequence} proposed an ensemble of Hidden Markov Models (HMM-e) to address challenges of class imbalance and fragmented behavioral data, showing that lightweight, interpretable sequence models can outperform or rival deep learning methods in tasks such as credit card fraud detection. Their findings reinforce the value of treating authentication attempts as sequences of actions rather than independent events, a perspective that directly informs our application of Bayesian approaches to optimize IVR authentication.

This project proposes a novel approach to optimize IVR authentication using Bayesian methods, which provide a probabilistic framework to dynamically update the belief about a caller’s legitimacy based on the outcomes of individual credential checks. By incorporating prior knowledge of fraud risk and analyzing conditional probabilities of pass/fail outcomes, we can design a more responsive and personalized authentication path. For instance, the system could select the next most informative credential based on previous results, maximizing the ability to confirm legitimate access or detect fraudulent behavior.

Ultimately, this research aims to develop a dynamic, risk-aware IVR authentication process that enhances fraud detection while preserving customer convenience—striking a more effective balance between security and usability.

\section{Methodology}

This study uses Bayes' Theorem and the Law of Total Probability to estimate the likelihood that an IVR caller is a fraudster, given their credential pass outcomes. The key idea is that fraud detection should not be a static, rule-based process, but a probabilistic one where every new piece of evidence (credential pass/fail) updates our belief about caller legitimacy. 

\subsection{Bayes' Theorem}
Let the event of interest A happen under any of the hypotheses \(H_i\) with a known (conditional)probability \(P(A|H_i)\) \cite{c7}. Assume that the probabilities of hypotheses \(H_1, ..., H_n\) are known (prior probabilities). Then the conditional (posterior) probability of the hypothesis \(H_i, \ i=1,2,...,n,\) given that event A happened, is 
\[P(H_i|A) = \frac{P(A|H_i)P(H_i)}{P(A)},\]
where 
\[P(A) = P(A|H_1)P(H_1)+...+P(A|H_n)P(H_n).\]

In our experiment, 

\[P(F|C=1) = \frac{P(C=1|F)P(F)}{P(C=1)}\]

where: 

\begin{itemize}
    \item \(P(F)\) is the prior probability of fraud (baseline fraud risk).
    \item \(P(C=1|F)\) is the likelihood that a fraudster can pass a credential. 
    \item \(P(C=1)\) is the overall probability of passing that credential. 
    \item \(P(F|C=1)\) is the posterior fraud probability after the credential is passed. 
\end{itemize}

For example, if the prior fraud rate is 3.88\% and a caller passes Credential A, which fraudsters pass 85.9\% of the time, the posterior fraud probability remains high at ~4.49\%. In contrast, if the caller also passes Credential E, which fraudsters rarely succeed on, the posterior drops sharply to 0.77\%. This illustrates how subsequent Bayesian updates quantify the real value of each credential in filtering out fraud.

\subsection{Law of Total Probability}

The Law of Total Probability is used to compute denominators such as \(P(C=1)\), ensuring that both fraudster and legitimate caller behavior are captured: 

\[P(C = 1) = P(C=1|F)P(F) + P(C=1|L)P(L)\]

Where L denotes a legitimate caller. This framework transforms authentication into a dynamic risk assessment process rather than a fixed checklist.

\section{Data Exploration}
\subsection{Dataset Introduction}
The dataset used in this study is a simulated fraud authentication dataset \cite{Data}that models caller verification in an Interactive Voice Response (IVR) customer service system. It contains 5,000 records and 11 columns, where 10 columns (A–J) represent credential checks and one column (is\_fraud) serves as the label.

Each credential column can take one of three values: 
\begin{itemize}
    \item 1: caller successfully passed the credential check
    \item 0: caller failed the check
    \item null: the credential is not available for the account
\end{itemize}

The label column is\_fraud indicates whether the caller is fraudulent (1) or legitimate (0). 

\subsection{Fraud Distribution}

The dataset is highly imbalanced. Out of 5,000 callers, only 194 (3.88\%) are labeled as fraudsters, while 4,806 (96.12\%) are legitimate callers. This imbalance reflects a realistic fraud detection environment, where fraudulent attempts are rare compared to legitimate customer interactions. 

\subsection{Missing Data Patterns}

Credentials vary in availability, simulating different customer profiles and authentication pathways:
\begin{itemize}
    \item Weak credentials such as A–G have relatively fewer missing values (ranging from 218 to 934 missing).
    \item Strong credentials H–J show substantial missingness, with over 40–60\% of values absent. For instance, credential J is missing in 3,006 cases.
    \item The label column is\_fraud has no missing values.
\end{itemize}

This missingness is consistent with how IVR systems selectively prompt for credentials depending on risk level, customer account setup, and requested service type.

\subsection{Correlation Between Credentials}
The current sequential authentication process in IVR systems is often designed under the assumption that each credential represents an independent checkpoint. This assumption implies that the probability of passing one credential is unrelated to the probability of passing another. In practice, however, this assumption does not hold. Many credentials exhibit conditional dependencies based on how users or fraudsters acquire or recall information. For example, a customer who can accurately provide their address information is also more likely to correctly provide their ZIP code, as both pieces of information may be sourced from the same document (e.g., a postcard or billing statement). In contrast, fraudsters may be able to guess or obtain weak credentials but are more likely to fail when challenged with less accessible ones. These interdependencies suggest that the outcome of one credential check can influence the likelihood of passing subsequent checks, particularly when considering differences between legitimate callers and fraudsters. The empirical analysis of the dataset reveals notable correlations among certain credentials.

\begin{figure}
    \centering
    \includegraphics[width=1\linewidth]{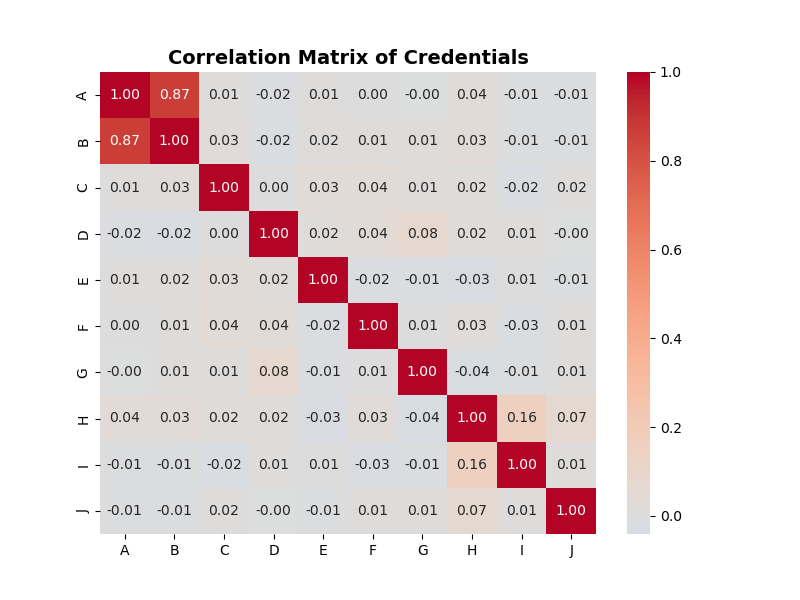}
    \caption{Correlation Between Credentials}
    \label{fig:correlation}
\end{figure}

As shown in the correlation matrix of credentials A–J \cref{fig:correlation}, most credential pairs exhibit near-zero correlation, indicating relative independence. However, some pairs stand out. For example, Credential A and Credential B have a correlation of 0.87, suggesting that they capture very similar caller information. This high degree of dependence undermines the value of treating them as separate authentication factors, because a fraudster who can successfully pass one is highly likely to pass the other as well. Moderate correlations also exist between Credential H and I (0.16) and Credential H and J (0.07), further illustrating that not all checkpoints provide independent security contributions.

The problem with assuming independence is that it can lead to overestimation of system security. If two correlated credentials are treated as independent, the combined fraud risk may be underestimated, creating a false sense of protection. For instance, relying on both A and B as a two-factor check may appear to strengthen authentication, but given their strong correlation, this effectively acts as a single factor from a risk perspective.

Therefore, when designing multi-factor authentication (MFA) strategies in IVR, it is essential to account for correlation between credentials. Using combinations of weakly correlated or independent credentials (e.g., A with E or H) provides stronger protection, as fraudsters face genuinely different verification barriers. Ignoring these correlations risks introducing redundancy into the process, while still leaving vulnerabilities that sophisticated attackers could exploit.

\subsection{Data Summary}

\begin{enumerate}
    \item Imbalance in fraud representation presents a challenge for model training and evaluation.
    \item Sequential credential checks mirror real-world authentication flows: some customers may only be prompted for a few credentials, while others undergo deeper verification depending on the service level.
    \item Data sparsity in strong credentials (H–J) suggests that not all features carry equal weight in caller verification. This provides a natural foundation for applying Bayes’ Theorem to dynamically adjust authentication strategies based on available evidence.
\end{enumerate}
\section{Fraud Authentication}

While the exact real-time pass or fail probabilities for fraudsters on individual credentials are not directly observable, they can be estimated using historical call logs, known fraud claims, or expert knowledge. The primary objective of this experiment is to establish a baseline for evaluating the likelihood of fraud when a fixed pair of credentials is successfully passed during IVR authentication. 

In this simulation, we model the IVR fraud authentication process after a caller is identified by credit card number. Following identification, the system attempts to authenticate the caller using up to ten available credentials associated with the customer’s profile. These credentials may include security questions, ZIP code, CVV, voice biometrics, or other identifying information.

\begin{figure}
    \centering
    \includegraphics[width=1\linewidth]{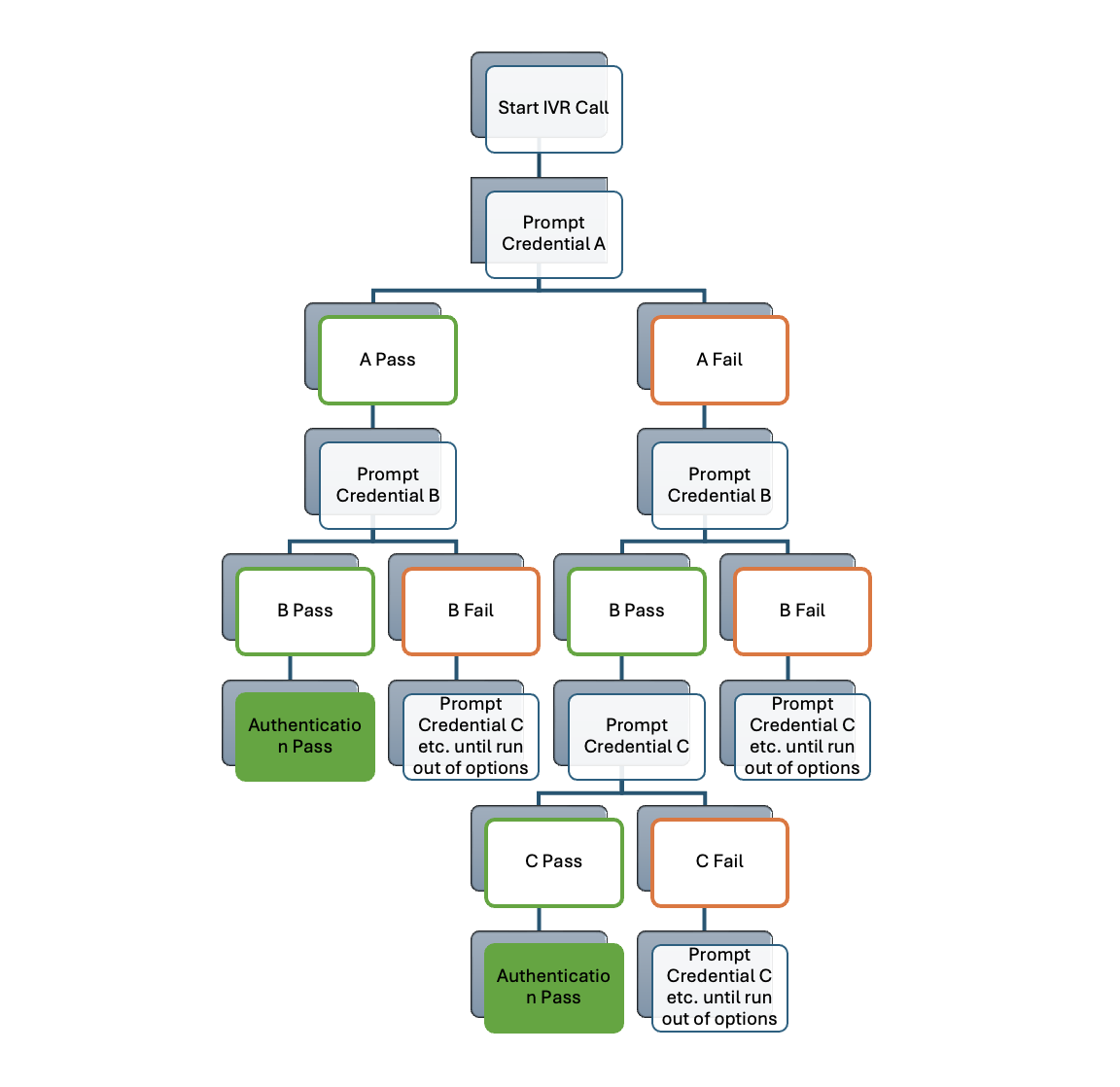}
    \caption{Sequential IVR Authentication Flow}
    \label{fig:Auth_flow}
\end{figure}

Under the current system design, a caller is granted access to services if they successfully pass at least two credentials, which are presented sequentially from A to J, see \cref{fig:Auth_flow} for demonstration. 

Based on these sample data, we calculated that the prior probability of fraud is:
\[P(Fraud) = 3.88\%\]

Although fraudsters cannot directly execute transactions through the IVR channel, they are able to exploit the system to harvest sensitive account information and transaction details. This harvested information can then be used to facilitate downstream fraudulent activity. By linking confirmed fraud transactions that occurred after IVR sessions back to their corresponding authentication attempts, we were able to establish a connection between IVR credential failures or successes and subsequent fraud outcomes. 

Empirical pass/fail probabilities for legitimate customers were derived from historical call data involving verified, non-fraudulent interactions, and are presented in \cref{tab:a1_cred_prob}. Furthermore, conditional fraud rates can be estimated given that a particular credential check is passed, providing insight into how individual credentials contribute to the overall authentication risk profile. 

\begin{table}
    \centering
    \begin{tabular}{|c|c|c|c|}
    \hline
     Credentials & Pass & Fail or Null & Fraud Rate when Pass\\
     \hline
A		&		0.8590		&		0.1410		&	4.494\%	\\
B		&		0.8086		&		0.1914		&	4.477\%	\\
C		&		0.7142		&		0.2858		&	3.500\%	\\
D		&		0.7866		&		0.2134		&	2.670\%	\\
E		&		0.6026		&		0.3974		&	0.664\%	\\
F		&		0.5474		&		0.4526		&	0.438\%	\\
G		&		0.4948		&		0.5052		&	0.000\%	\\
H		&		0.2966		&		0.7034		&	0.067\%	\\
I		&		0.1944		&		0.8056		&	0.000\%	\\
J		&		0.1124		&		0.8876		&	0.000\%	\\
     \hline
    \end{tabular}
    \caption{Credential Pass Probabilities for IVR Callers}
    \label{tab:a1_cred_prob}
\end{table}

\section{Multi-factor Authentication}

In IVR systems, Multi-factor Authentication (MFA) is a security mechanism that requires callers to provide two or more independent credentials before gaining access to services. Unlike single-factor authentication, which relies on only one piece of information (such as a PIN or account number), MFA leverages a layered approach that makes unauthorized access significantly more difficult for fraudsters. Each additional credential acts as an independent checkpoint, thereby reducing the overall likelihood that an attacker can successfully impersonate a legitimate customer.

An important consideration in designing MFA for IVR is that not all credentials provide equal protection. For example, credential A has a pass rate of 85.9\%, but the fraud rate among callers who pass A remains relatively high at 4.49\%. This indicates that credential A on its own is relatively weak, as it can be more easily obtained or guessed by fraudsters. To strengthen the authentication process, it is therefore necessary to combine A with additional credentials, which makes the probability of fraudster success conditional on passing multiple independent checks.

When we extend the verification process beyond a single credential, the benefits of MFA become evident. As shown in \cref{tab:mfa_example}, pairing credential A with other checks substantially lowers the fraud rate. For instance, when credential B is prompted after A, the fraud rate remains at 4.52\%, providing limited improvement. In contrast, combining credential A with credential E reduces the fraud rate to 0.77\%, while combinations such as A with F, G, I, or J result in near-zero fraud rates. These results demonstrate how careful selection of complementary factors in MFA can dramatically enhance security in IVR authentication, even when one of the credentials is relatively weak.

The key insight is that MFA not only strengthens the defense against account takeovers but also allows for risk-based optimization of the authentication flow. By applying empirical fraud rate analysis and Bayesian reasoning, IVR systems can dynamically select the most effective set of credentials given the caller’s profile and the service requested. This ensures that fraudsters face increasingly difficult barriers while minimizing friction for legitimate customers, striking a balance between security and usability.

\begin{table}
    \centering
    \begin{tabular}{|c|c|}
    \hline
    Combination of Two Credential & Fraud Rate \\
    \hline
P(A==1 \& B==1)	&	4.519\%	\\
P(A==1 \& C==1)	&	4.038\%	\\
P(A==1 \& D==1)	&	3.091\%	\\
P(A==1 \& E==1)	&	0.771\%	\\
P(A==1 \& F==1)	&	0.513\%	\\
P(A==1 \& G==1)	&	0.000\%	\\
P(A==1 \& H==1)	&	0.078\%	\\
P(A==1 \& I==1)	&	0.000\%	\\
P(A==1 \& J==1)	&	0.000\%	\\
\hline
    \end{tabular}
    \caption{Multi-factor Authentication Example}
    \label{tab:mfa_example}
\end{table}

\section{Adaptive Credential Ordering to Improve Fraud Detection}

Fraudsters possess varying levels of customer information, enabling them to selectively bypass stronger authentication credentials while targeting weaker ones that are easier to obtain, often through data breaches or social engineering. For instance, fraudsters are most likely to attempt credentials A and B, since the pass rate of these two credentials are \(85.90\%\) and \(80.86\%\) respectively, and least likely to attempt credential J which only has pass rate of (\(11.24\%\)). This empirical distribution highlights that not all credentials provide equal deterrence value; certain credentials inherently serve as stronger barriers against fraudulent access. 

Traditional IVR systems prompt credentials in a fixed order, ignoring this behavioral variation. Our approach uses Bayes’ Theorem to adaptively sequence credentials. After each credential outcome, the posterior fraud probability is updated, and the system selects the next most informative credential—the one expected to reduce fraud risk the most while keeping legitimate customer friction acceptable.

Formally, if a caller passes credential \(C_i\), the updated fraud probability is:

\[P(F|C_i = 1) =\frac{P(C_i = 1|F) \cdot P(F)}{P(C_i= 1)}\]

At the next step, the system evaluates which candidate credential \(C_j\) minimizes the expected posterior fraud probability \(P(F|C_i =1, C_j = 1)\). 

For example, pairing A with B provides little additional value since they are highly correlated: fraud probability only drops from 4.49\% to 4.52\%. By contrast, pairing A with E reduced posterior fraud risk to 0.77\%, and A with G or I results in near-zero fraud probability. The lesson is that effective ordering depends not only on individual credential strength, but also on how independent two credentials are from each other. 

This adaptive design also accommodates operational trade-offs. Some pairs, such as A+G, block 100\% of fraud but also reject over 55\% of legitimate customers—too costly for everyday use. However, they can serve as high-risk paths, applied only when prior fraud risk is already elevated. Thus, Bayesian ordering balances three goals: deterrence, risk reduction, and customer experience.

\subsection{Algorithmic Approach}

We build on the fusion framework of Panigrahi et al. (2009), who combined rule-based filters, Dempster–Shafer theory, and Bayesian learning to update suspicion scores in fraud detection. In their system, initial beliefs are derived from multiple evidences, then revised using Bayes’ rule to identify the maximum a posteriori hypothesis and guide threshold-based decisions \cite{Panigrahi}. We adapt this approach to the IVR context by treating credential attempts and call-risk signals as evidences, updating posterior beliefs after each event to optimize credential sequencing and reduce fraudster success.

The proposed framework evaluates each possible two-step credential ordering. The ideal first credential is one that maximizes deterrence (i.e., has the lowest fraudster attempt probability), while the ideal second credential is chosen to minimize fraud probability conditional on the first being passed. 

\begin{enumerate}
    \item Estimate base rates: Using historical IVR call and fraud investigation records, estimate the prior fraud probability
\(P(F)\) for incoming calls.
    \item Model credential attempt behavior: For each credential \(C_i\), estimate:
        \begin{itemize}
            \item \(P(pass|F)\): probability that a fraudster can pass this credential. 
            \item \(P(pass|L)\): probability that a legitimate customer passes this credential, where L denotes "legitimate".
        \end{itemize}
    \item Bayes' updating: When a caller attempts credential \(C_i\):
        \begin{itemize}
            \item If they pass, update the posterior fraud probability using Bayes' Theorem.
            \item If they fail, the system can block or escalate, depending on policy. 
        \end{itemize}
    \item Dynamic credential ordering: At each step, select the next credential that maximize fraud detection gain. A natural criterion is to minimize the expected posterior fraud probability conditional on a pass, while also controlling for the legitimate customer failure rate. 
    \item Stopping rules: If the posterior fraud probability falls below a safety threshold (e.g., <0.1\%), the system can end authentication early and allow access. Conversely, if the posterior exceeds a risk threshold (e.g., >50\%), the call is blocked or escalated to a fraud specialist.

\end{enumerate}

\subsection{Experiment}

We assume a caller must pass both credentials in a pair to be allowed forward. If they fail either, they’re blocked. Some measurements are consider: 
\begin{itemize}
    \item Fraud block rate (TPR): fraction of fraud calls that would be blocked by failing at least one of credentials. Higher is better.
    \item Legit block rate (FPR): fraction of legitimate calls that would be blocked. Lower is better, since this will negatively impact legitimate customers. 
    \item Posterior fraud risk if passed: \(P(Fraud | C_i=1 \& C_j=1)\) Evaluate fraud risk if both credentials are passed.
    \item Youden’s J: \(TPR - FPR\). A simple balance of catch fraud vs. customer friction; higher is better.
    \item Pass-both rate: share of all calls that would pass both (capacity/usability signal).
\end{itemize}

In the experiment, we based on fraud prevention logic and use posterior risk to pick pairs that minimize \(P(Fraud | C_i=1 \& C_j=1)\). And compare metrics for all \({10}\choose{2}\) = 45 pairs and rank them. 

Among callers who pass these pairs, the observed fraud rate is \(0\%\) for combinations of \(A + G\) and \(B+ G\), which is great from a "confidence-after-pass" perspective. They also block all fraud in the sample data, but with non trivial legit customer friction \(~ 55.7-58.3\% \) FPR, \cref{tab:cred_comb}. If the operation center can absorb that friction or use these as a high-risk path, they are strong "safe-after-pass" gates.

\begin{table}
    \centering
    \begin{tabular}{cccccc}
    \hline
        Pairs & Fraud Rate & TPR & FPR & Pass-both & Youden's J\\
        \hline
     A + G    & 0.000 & 1.000 & 0.557 & 0.426 & 0.443 \\
     B + G    & 0.000 & 1.000 & 0.583 & 0.401 & 0.417\\
     \hline
    \end{tabular}
    \caption{Credential Combinations Minimize the Fraud Risk}
    \label{tab:cred_comb}
\end{table}

\section{CONCLUSIONS}

This research demonstrates how Bayesian inference can transform IVR authentication from a static checklist into a dynamic, risk-aware process.

\begin{itemize}
    \item In static analysis, we found that some credential pairs (e.g., A+B) leave fraud rates virtually unchanged \(~4.5\%\), while others \(A+E\) reduce fraud risk sixfold, from \(4.49\%\) to \(0.77\%\).
    \item In deterrence focused experiments, leading with stronger credentials (e.g., D) discouraged fraudster attempts more effectively than starting with weaker ones. 
    \item In adaptive sequencing, dynamically selecting credentials based on previous outcomes further reduced posterior fraud probabilities, with pairs like A→I or A→G achieving 0\% observed fraud among passers.
\end{itemize}

Together, these results show that even small adjustments in credential ordering can yield disproportionately large gains in fraud detection. By combining empirical pass/fail data with Bayesian updating, banks can implement authentication flows that not only block fraudsters but also minimize unnecessary friction for legitimate customers.

Adopting such adaptive strategies provides a path toward IVR systems that are not just automated, but intelligently risk-aware—offering stronger protection against evolving fraud threats while preserving the customer experience.

\section{Discussion and Future Research}

While our experiments highlight effective strategies for minimizing fraud risk in IVR authentication, several areas remain for future exploration. First, this study assumes access to reasonably accurate estimates of credential-level pass rates and conditional probabilities, which may be difficult to obtain in practice. Future work could explore methods to dynamically learn these probabilities from real-time call data using Bayesian updating or machine learning techniques. Additionally, while we focused on two-step authentication paths, extending this framework to multi-step adaptive flows (e.g., selecting the best sequence of three or more credentials) could further improve fraud detection with minimal customer friction.

Another important direction involves modeling legitimate customer behavior more fully, including failure rates, retry patterns, and abandonment tendencies. Integrating this with fraudster behavior could support a more comprehensive cost-benefit framework, balancing fraud loss prevention with customer experience. Finally, incorporating contextual signals—such as call velocity, ANI mismatch, and time-of-day patterns—into the Bayesian model may offer richer, real-time fraud risk scoring. These extensions would help transform static IVR authentication into a truly intelligent, adaptive fraud prevention system.

\addtolength{\textheight}{-12cm}   % This command serves to balance the column lengths

%%%%%%%%%%%%%%%%%%%%%%%%%%%%%%%%%%%%%%%%%%%%%%%%%%%%%%%%%%%%%%%%%%%%%%%%%%%%%%%%

%%%%%%%%%%%%%%%%%%%%%%%%%%%%%%%%%%%%%%%%%%%%%%%%%%%%%%%%%%%%%%%%%%%%%%%%%%%%%%%%

%%%%%%%%%%%%%%%%%%%%%%%%%%%%%%%%%%%%%%%%%%%%%%%%%%%%%%%%%%%%%%%%%%%%%%%%%%%%%%%%
\bibliographystyle{plain}

\bibliography{refs}

\end{document}